\documentstyle [12pt,aps,prl]{revtex}
\begin{document}
\title{Excitonic instability and electric-field-induced phase
          transition towards a two dimensional exciton condensate}
\author{ 
   \\ Y. Naveh\cite{email Naveh} 
and B. Laikhtman\cite{email Laikhtman}
\\ 
   {\normalsize \it Racah Institute of Physics, The Hebrew
        University,} \\
   {\normalsize \it Jerusalem 91904, Israel } } 
\maketitle

\begin{abstract} 

We present an InAs-GaSb-based system in which the electric-field 
tunability of its 2D energy gap implies a transition towards a
thermodynamically stable excitonic condensed phase. Detailed
calculations show a 3 meV BCS-like gap appearing in a second-order
phase transition with electric field. We find this transition to be
very sharp, solely due to exchange interaction, and so, the exciton
binding energy is greatly renormalized even at small condensate
densities. This density gradually increases with external field, thus
enabling the direct probe of the Bose-Einstein to BCS crossover.

\end{abstract}

The long search for a condensed phase of  excitons has greatly
expanded in recent years\cite{Griffin 95}. Although theoretically
recognised three decades ago\cite{Keldysh 65}, experimental
indications of the existence of such a phase emerged only recently. In
${\rm Cu_2O}$ bulk samples, The time-evolution 
of the excitonic line-shape\cite{Snoke 87}, as
well as macroscopic ballistic transport of excitons\cite{Fortin 93},
are the most remarkable manifestations of the appearance of an exciton
condensed phase. Indications of a two-dimensional (2D) condensate
under strong magnetic field in semiconductor heterostructures were
also reported\cite{Fukuzawa 90,Cheng 95}.

Two major obstacles interfere in forming an exciton condensate. One is
the short recombination lifetime ($\sim 1 \, {\rm ns}$) of
photo-excited direct excitons. This time can be shorter than the time
free excitons need to thermalize and condense. The recombination rate
also produces enough heat to possibly destroy the condensate. In ${\rm
Cu_2O}$, its unique crystal structure results in a dipole-forbidden
recombination time of $\sim 10 \, \mu {\rm s}$.  A usual way to
increase the exciton lifetime in 2D structures is to introduce a wide
barrier material between separated electron and hole
materials\cite{Fukuzawa 90}. This procedure, however, also reduces the
electron-hole (e-h) Coulombic interaction, and so decreases $T_c$, the
critical temperature for condensation.

The second obstacle is the usually more favorable formation of an
e-h droplet. Although that liquid phase may have a non-zero
condensate order parameter, this is usually suppressed by the
interactions leading to the formation of the droplet. It was
suggested\cite{Korolev 94} that a strong magnetic field should reduce
the interaction between excitons, thus favoring their quantum 
condensation. Sophisticated heterostructure design has also been
proposed\cite{Zhu 95} to overcome the droplet formation.

Both of this obstacles are immediately eliminated if the thermodynamic
equilibrium state of the system is a condensed excitonic phase. Such
is the case in a system with an excitonic instability\cite{Kozlov
65,Jerome 67}, namely a system with a semiconducting-like band
structure, but with a single-particle energy gap which is smaller than
the exciton binding energy. Excitons now form spontaneously, and the
recombination lifetime is infinite. Compared to the unstable
semiconducting ("vacuum") state, the excitonic state has negative
energy, while a state of free e-h plasma has a positive
energy. Therefore, at temperatures lower than the original energy gap,
even if an e-h droplet forms, it must have a non-zero order parameter,
and the excitonic phase exists even at high densities. 

If, in addition to being unstable, the system allows external control
over the original energy gap, then another important feature arises:
The density of the exciton condensate can be externally controlled, and
different density-regimes can thus be probed in an otherwise identical
system.

In this paper we present a realistic 2D system which exhibits an
excitonic instability. The specific structure we propose can be
readily fabricated by existing epitaxial techniques. Its 2D band
structure is very sensitive to externally applied electric fields. In
particular, by applying electric field, one can continuously open an
energy gap ranging from zero to a few meV. The negative (semimetallic)
case is not discussed here. 

We rigorously study the properties of the system using the Keldysh-BCS
theory, described later. Starting from the large-gap state, we find an
excitonic-condensate phase transition which occurs exactly when the
gap reaches the value of the exciton binding energy
$\varepsilon_0$. closing the gap further results in an increase of the
condensate density $n$, up to a density of $nd_0^2 \sim 4$ at zero
energy gap. $d_0 = 2 \hbar / \sqrt{2m\varepsilon_0}$ is twice the exciton
Bohr radius ($m$ is the reduced effective mass). 
In the theory we use zero temperature, and isotropic band
structures of the underlying materials.  However, in view of the
instability-origin of the condensed phase, finite temperature,
warping, and imperfections should not change the results significantly
(as long as the temperature is well below the energy of the original
gap).

A prototype of the studied heterosystem was described in ref.\
\cite{Naveh 95}. Its energy profile is shown in fig.\ 1. In this
staggered-gap system, the valence band edge in the GaSb layer lies
above the conduction band edge of the InAs layer. Due to the strong
mismatch between valence and conduction states, a thin ({\it i.e.},
2-4 atomic layers) AlSb barrier is enough to suppress  any e-h
coupling between the active  materials\cite{Naveh unpb}, and we have a
regular semimetallic parabolic band structure. such a thin barrier,
however,  has a negligible effect on the interlayer coulomb
interaction. We designate this case as case (I). Without an AlSb
interlayer, there exists quantum mechanical coupling between the two
layers\cite{Naveh 94}. The band structure of the system is then
composed of 2 W-shaped bands\cite{Naveh 95}, and the particles at the
band edges are of mixed type. This is case (II). 

In both cases, small electric fields applied across the
heterostructure greatly change its band structure. In case (I) the
band overlap decreases with field, until a zero gap is reached, beyond
which a semiconductor energy gap opens. This gap now increases with
applied field at a rate of $1 \,\rm{meV} \rm{cm} / \rm{KV}$. In case
(II), the band structure is an indirect-gap-semiconductor one, with
the main effect of the field being the shift of the band extrema
towards larger wavenumbers. In this case an unticrossing energy gap of
value 2-5 meV is present at any field. However, it is possible to
reduce the value of the gap by means of a very narrow AlSb barrier, or
any other degradation of the GaSb-InAs interface. These results  are
summarized  in  insets (a) and (b) of Fig.\ 1. It is important to note
that the total charge in the system can be independently controlled by
means of a double-gated structure with in-plane leads\cite{Naveh 95},
so we are allowed to consider a neutral sample even if the structure
is originally unintentionally doped.

An excitonic instability happens in cases (I), (II) when the gap is
smaller than the exciton binding energy $\varepsilon_0$. In case (II)
the particles at the band edges are of mixed type, so $\varepsilon_0$
is much reduced. We therefore treat in this work only case (I), where
no coupling is allowed. The formal way to treat the instability is by
renormalization of the band structure\cite{Jerome 67,Keldysh 68,Comte
82}. This amounts to a BCS-like procedure, where the electrons and
holes play the role of opposite-spin particles in the regular BCS
theory. The outcome of the theory can be summarized in terms of
$\Delta(k)$, the BCS gap function, which describes the band
renormalization, and $n$, the condensate density. When $nd_0^2 \ll 1$,
$n$ is actually the exciton density, and the condensed phase is a
Bose-Einstein condensate (BEC) of the Kosterlitz-Thouless type. In the
opposite limit, $n$ designates the density of "Cooper pairs", which
are zero momentum e-h complexes in the Cooper-pairing
sense. The transition between the two regimes is believed to be
continuous\cite{Comte 82,Randeria 95}. We suggest the present system
as an experimental platform to examine this BEC to BCS crossover.

To formulate the theory we start with the e-h Hamiltonian
\begin{eqnarray}	\label{hamiltonian}
   H & = & \sum_{\bf k} \left( 
	\varepsilon_{\bf k}^e a_{\bf k}^\dagger a_{\bf k} + 
	\varepsilon_{\bf k}^h b_{\bf k}^\dagger b_{\bf k} \right) + \\
     &	 & \frac{1}{2D^2} \sum_{\bf k,k',q} \left(      
	V_q^{ee} a_{\bf k}^\dagger a_{\bf k'}^\dagger 
	a_{{\bf k'}+{\bf q}} a_{{\bf k}-{\bf q}} + 
	V_q^{hh} b_{\bf k}^\dagger b_{\bf k'}^\dagger 
	b_{{\bf k'}+{\bf q}} b_{{\bf k}-{\bf q}} - 
	2 V_q^{eh} a_{\bf k}^\dagger b_{\bf k'}^\dagger 
	b_{{\bf k'}+{\bf q}} a_{{\bf k}-{\bf q}} \right)  \nonumber
\end{eqnarray}
where $\varepsilon_{\bf k}^{e,h}$ are the electron and hole energies
(before renormalization), measured from their respective band edges,
 $V_q^{\alpha\beta} = 2\pi e^2 / (\epsilon q) \,
F^{\alpha\beta}(q)$, 
$\epsilon$ the dielectric constant, and $D^2$ is the sample area.
$a_{\bf k}$ and $b_{\bf k}$ are electron and hole annihilation
operators. The structure factors $F^{\alpha\beta}(q)$ are
found to be:
\begin{mathletters} \label{struct}
\begin{eqnarray}
   F^{\alpha\alpha}(q) & = & \frac{8\pi^2}{\zeta_\alpha (4\pi^2 + 
	\zeta_\alpha^2)} \left[ 1 + \frac{3\zeta_\alpha^2}{8\pi^2}
	- \frac{4\pi^2}{(4\pi^2 + \zeta_\alpha^2)\zeta_\alpha}
	\left( 1 - e^{-\zeta_\alpha} \right) \right] \\
   F^{eh}(q) & = & \frac{16\pi^4}{\zeta_e \zeta_h} 
	\frac{\left( 1 - e^{-\zeta_e} \right) 
	\left( 1 - e^{-\zeta_h} \right)}{
	(\zeta_e^2 + 4\pi^2) (\zeta_h^2 + 4\pi^2)}, 
\end{eqnarray}
\end{mathletters}
where $\zeta_{e,h} = qL_{e,h}$, $L_{e,h}$ being the widths of the
wells. It is worth mentioning that the structure factors
reduce the binding energy, as well as the BCS gap, by more than a
factor of three, as compared to the truly 2D case.

As usual\cite{Jerome 67,Keldysh 68,Comte 82}, we proceed by making a
Bogoliubov transformation of $H$, which results with a numerical (not
operator) term in the Hamiltonian. Minimizing this term with respect
to the transformation constant leads to the generalized BCS
equations:
\begin{mathletters} \label{bcs}
\begin{eqnarray}
   \xi_{\bf k} & = & \varepsilon_{\bf k} +E_g - 
	\int V^{pp}_{{\bf k}-{\bf k'}} 
      \left( 1-\frac{\xi_{\bf k'}}{E_{\bf k'}} \right) \, 
	\frac{d^2k'}{(2\pi)^2}
      \label{bcsa} \\
   \Delta_{\bf k} & = & \int V^{eh}_{{\bf k}-{\bf k'}}
      \frac{\Delta_{\bf k'}}{E_{\bf k'}} \, 
       \frac{d^2k'}{(2\pi)^2}  \label{bcsb} \\
   E_{\bf k} & = & \left( \xi_{\bf k}^2 + \Delta_{\bf k}^2
      \right)^\frac{1}{2}.
							\label{bcsc}
\end{eqnarray}
\end{mathletters}
Here $\varepsilon_{\bf k} = \varepsilon_{\bf k}^e + \varepsilon_{\bf
k}^h$, $V^{pp} = \frac{1}{2} (V^{ee} + V^{hh})$, 
and $E_g$ is the energy gap.

 The integral term in Eq.\ (\ref{bcsa}) results from the
particle-particle Fock energy.  When this term is set to zero, the
system (\ref{bcs}) becomes the regular BCS gap equation. We show later
that the Fock term qualitatively alters the results. In
particular, due to this term, the transition to the excitonic phase is
very sharp.

The outcome of the self-consistent equations (\ref{bcs}) are the BCS
gap function $\Delta_{\bf k}$, which determines the renormalized
spectrum $E_{\bf k}$, and the condensate density\cite{Keldysh 68}
\begin{equation} \label{density} 
   n = \frac{1}{2 \pi^2} \int \left(
   1-\frac{\xi_{\bf k}}{E_{\bf k}} \right) \, d^2k,
\end{equation} 
which includes here spin degrees of freedom.
Eqs.\ (\ref{bcs}) can be solved analytically in the limits of large
and small densities\cite{Jerome 67,Comte 82}. The onset of the
condensed state is found by solving Eqs.\ (\ref{bcs}) in the first
approximation in $\psi_{\bf k} \equiv \frac{\Delta_{\bf k}}{E_{\bf
k}}$, where they become: 
\begin{equation}        \label{schrodinger}
    \varepsilon_{\bf k} \psi_{\bf k} - \int V^{eh}_{|{\bf k}-{\bf k'}|} 
    \psi_{{\bf k'}} \, \frac{d^2k'}{(2\pi)^2} = -E_g \psi_{\bf k}.	
\end{equation} 
This is exactly the schr\"{o}dinger equation for the excitonic
problem. It is therefore concluded that the condensed state starts to
form precisely when the gap $E_g$ equals the exciton binding energy,
regardless of the shape of the spectrum, and the finite width of the
wells. 
                                           
To solve the problem completely, we define  $x_{\bf k} \equiv \tan
\varphi_{\bf k} \equiv  \frac{\xi_{\bf k}}{\Delta_{\bf k}}$ to obtain
a single equation
\begin{equation} \label{single}
   x_{\bf k} \int V^{eh}_{|{\bf k}-{\bf k'}|}
     \frac{1}{\sqrt{1+x_{\bf k'}^2}}
     \, \frac{d^2k'}{(2\pi)^2} = \varepsilon_{\bf k} +E_g - 
	\int V^{pp}_{|{\bf k}-{\bf k'}|}
     \left( 1 - \frac{x_{\bf k'}}{\sqrt{1+x_{\bf k'}}} \right)
     \frac{d^2k'}{(2\pi)^2}.
\end{equation}

By solving Eq.\ (\ref{single}) we analyze the excitonic condensate as a
function of the original energy gap of the unstable-spectrum.

Results for $n$ and $\Delta$ are shown in Figs.\ 2 and 3. The energy
gap is easily controlled experimentally by means of a perpendicularly
applied voltage. Starting from large gaps we reach a second order
phase transition when $E_g = \varepsilon_0$. Below $\varepsilon_0$,
the condensate density grows gradually towards its value of $nd_0^2 =
3.8$ at $E_g = 0$. The results near $E_g = 0$ are of less
significance, due to the fact that small deviation from $T=0$ would
change the picture. Screening effects are also expected to become
important for $nd_0^2 \gg 1$. However, we see that for $E_g > 0.5 \,
{\rm meV}$, where low finite temperatures should not have an
effect, all density regimes are present.

The same phase transition is also seen by considering the gap function
$\Delta$. Unlike the BCS behavior, $\Delta(k)$ is a nonmonotonic
function of the wavenumber $k$. This is solely due to the exchange
interaction term, which acts effectively as negative kinetic energy.
When exchange is not neglected, $\xi_k$ assumes negative values at
small wavenumbers, and the maximum of $\Delta(k)$ occurs roughly  when
$\xi_k$ crosses zero.

The maximum value of $\Delta(k)$, $\Delta_{\rm max}$, is a measure of
the strength of the condensate. Its value, about 3 meV, indicates that
the condensate should be readily observed at helium temperatures in
fair-mobility samples. The Fock term in Eqs.\ \ref{bcs} is again
responsible for the nonmonotonic dependence of $\Delta_{\rm max}$ on
$E_g$. It can be understood if negative values of $\xi_k$ are
considered as corresponding to single-particle semimetallic spectrum.
Then, reducing $E_g$ is equivalent to increasing the semimetallic
overlap, which is known\cite{Kozlov 65} to reduce $\Delta$.

The most obvious way to experimentally identify the formation of the
condensate is to measure the dependence of the optical absorbtion edge
on electric field. This can be done directly, or by measuring the
typical temperature for the temperature-activated conductivity of the
system. Starting from a wide-gap structure, the edge, which is at
$E_{\rm min}$, the minimum value of $E_k$, is located at an energy
$E_g$, with a separated excitonic peak at $E_g - \varepsilon_0$.
Closing the gap by application of field shifts these signals to lower
energies, and diminish the exciton peak, until it totally vanishes
when $E_g = \varepsilon_0$. Beyond this point, The absorbtion edge
steeply increases with decreasing field, and that is identified with
the formation of the condensate. The fact that $E_{\rm min}$
dramatically increases for $E_g < \varepsilon_0$, while the excitons
are still well separated, $nd_0^2 \ll 1$, means that the exciton
binding energy is greatly enhanced when the excitons form a BEC. This
renormalization of $\varepsilon_0$ is attributed to the interaction
between excitons, which is repulsive, and comes about through the
particle-particle exchange term. Although this interaction can be
small for small densities, it greatly affects the binding energy in a
ground-state condensate, where the kinetic energy of each particle is
zero. The effect is emphasized by the fact that $V^{pp}_{\bf kk'} \gg
V^{eh}_{\bf kk'}$, the latter which governs the internal binding
energy of the exciton. In contrast, if the exchange term is not
included in the BCS equations, then the exciton binding energy (and
$E_{\rm min}$) remain constant until $nd_0^2 \sim 1$.

The results provide a strong motivation to study the proposed material
in the goal of achieving a strong, well defined, stable, condensed
phase of excitons. Unlike other proposed, or experimentally studied
structures, here we present a condensate which is inherent to the
structure, and constitutes its equilibrium ground state even at
temperatures as high as 0.5 meV. The density of the condensate can be
easily changed here by means of electric field. Everything else
being unchanged, the BEC to BCS crossover can thus be readily studied. 
Two previous works concerned excitonic phases in InAs-GaSb related
structures. In \cite{Cheng 95} experiments were done in  semimetallic
conditions under strong magnetic fields. In \cite{Zhu 90}
single-exciton binding energies were calculated for different well
widths. Such calculations cannot account for the condensate even at
very low densities, where, as we have shown, many-body effects are of
extreme importance. The effect of electric field on the condensate in
these staggered-gap structures, and the externally induced phase
transition, is recognized here for the first time.

Delicate questions concern the problem of whether the exciton
condensate is superfluid. As is well known, macroscopic condensation
in the ground state is not sufficient for superfluidity.
Degeneracy of the quantum mechanical phase is also necessary. Indeed,
It was argued\cite{Jerome 67,Guseinov 73} that the exciton condensate
cannot be superfluid. The arguments are based on the suppression of
superfluidity by interband transitions. Unanswered
questions remain about the superfluid behavior in the limit when the
matrix elements for those transitions
are very small. Experiments in ${\rm Cu_2O}$ exhibit macroscopic
ballistic transport of excitons\cite{Fortin 93}, which can be interpreted
as the result of superfluidity. Lozovik and Yudson\cite{Lozovik 76}, and 
Shevchenko\cite{Shevchenko 77} 
showed that if the electrons and holes are
separated, the condensate can become superfluid (in that case it
actually becomes a double-layer 'capacitor-superconductor').
Shevchenko recently presented a phase diagram for that 
case\cite{Shevchenko 94}. In the presently studied structure, electrons
and holes are separated in type (I) structures. In type (II) structures
the low-energy particles are of mixed type, and are spread across the
double layer. Based on existing knowledge, one expects to
experimentally find
different superfluid behavior between the two structures, and so base
the ideas behind the suppression of superfluidity on firm grounds.

In conclusion, we presented a prototype system which we believe should
exhibit a condensed phase of excitons. This phase results from an
instability of the small-gap band structure, and is the thermodynamic
equilibrium state of the system. It is therefore expected to be robust
to finite temperatures and imperfections. The parameters of the
condensate, and of the underlying band structure, are easily
controllable by virtue of applied electric field. We thus propose the
material as a new platform for studying the condensed phase of
excitons.

Discussions with E. E. Mendez, J. K. Jain and P. Thomas are gratefully
acknowledged. This research was supported by the Israel Science
Foundation administered by the Israel Academy of Sciences and
Humanities. 

\newpage
\section* {Figure captions}
   
Fig.\ 1.  Relevant band edges of the studied heterostructure. $v$, $c$
denote valence and conduction band edges. $E_{v1}$, $E_{c1}$ are the
first confined levels. Layer widths correspond to 8 monolayers GaSb
and 28 monolayers InAs. Inset (a) shows the in-plane spectrum of the
carriers in the structure at zero electric field (solid lines). 
Dashed lines: spectrum of coupled particles in a structure without the
narrow AlSb barrier. Inset (b) shows the electric-field-dependence of
the energy gap $E_{c1} - E_{v1}$  for both cases.

Fig.\ 2.  Density of the condensate as a function of the original (not
renormalized) energy gap. This gap scales linearly with external
field.

Fig.\ 3.  Dependence of the condensate's parameters on the original
gap. Solid lines: maximum of $\Delta_k$. Dashed lines: Minimum of
$E_k$ ({\it i.e.}, the absorbtion edge). The inset shows $\Delta(k)$
for zero gap. "Complete" and "BCS" correspond, respectively, to
calculations including, or excluding, the Fock term in Eq.\
(\ref{bcsa}).


\begin{thebibliography}{99}
\bibitem{Naveh unpb} This total suppression is due also to the height
of the AlSb barrier, and can be shown by any microscopic ({\it e.g.},
tight binding) model.
\end{thebibliography}
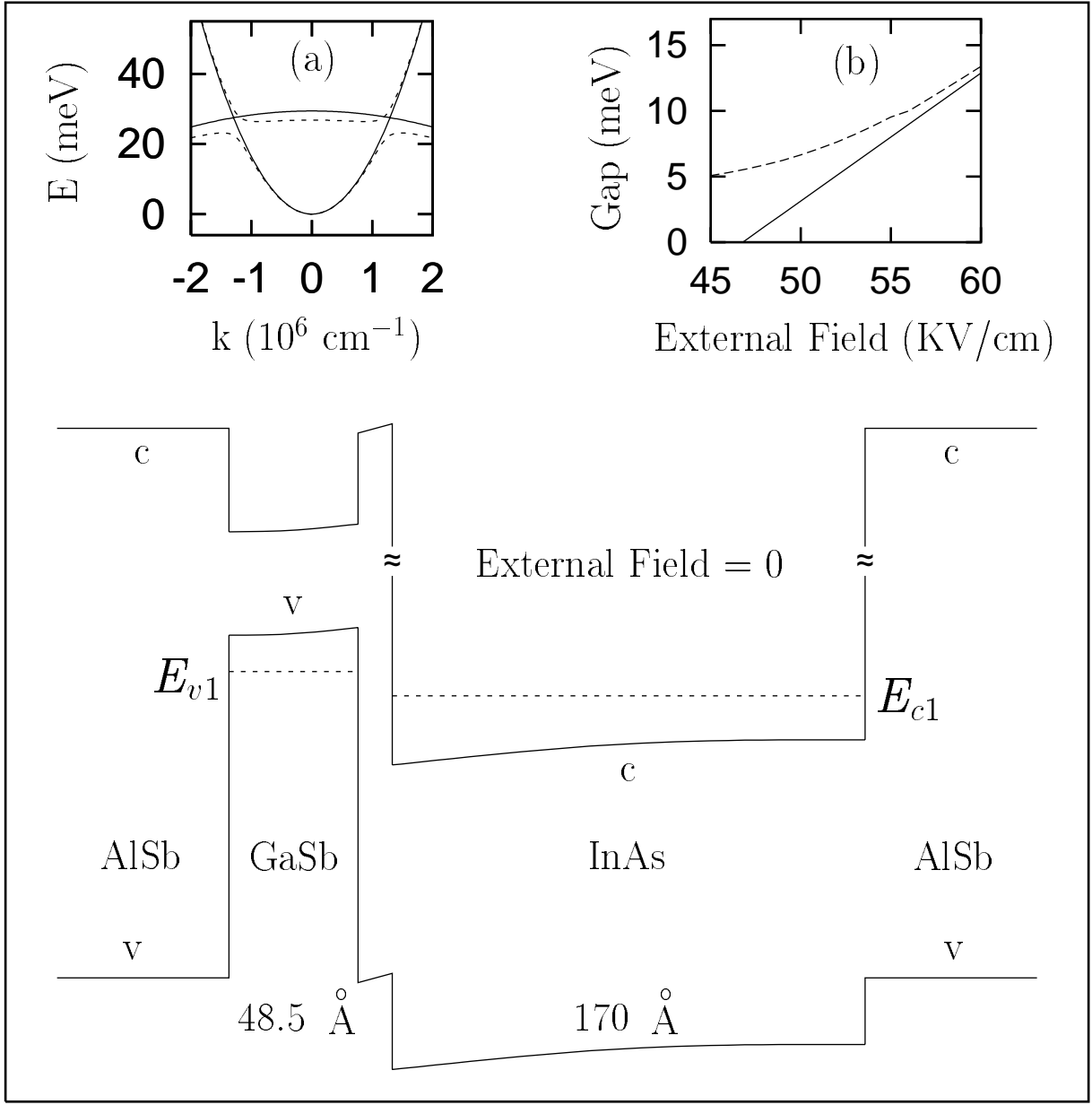
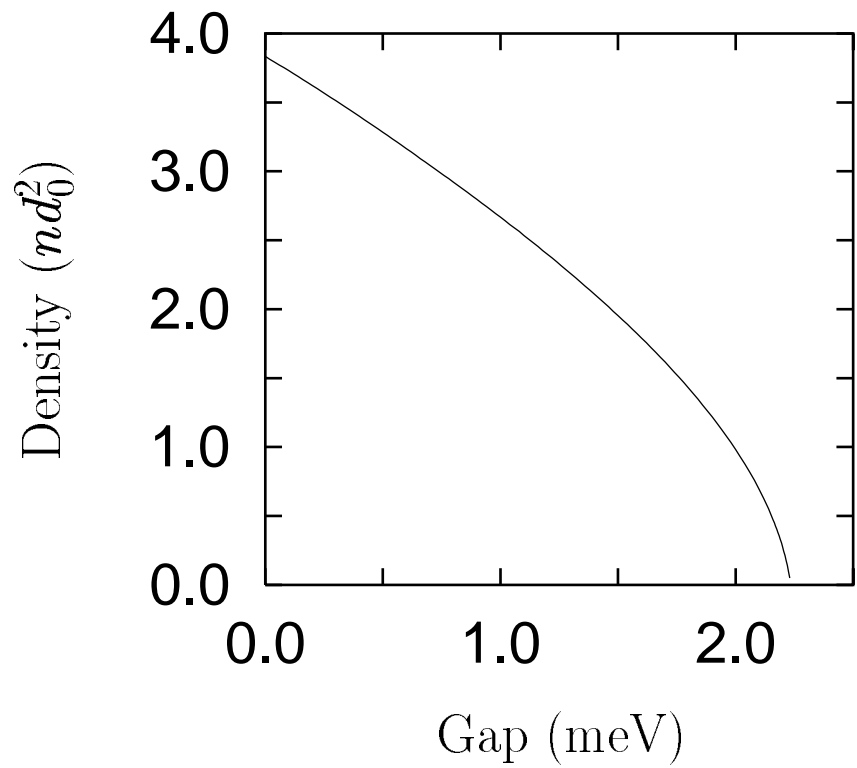
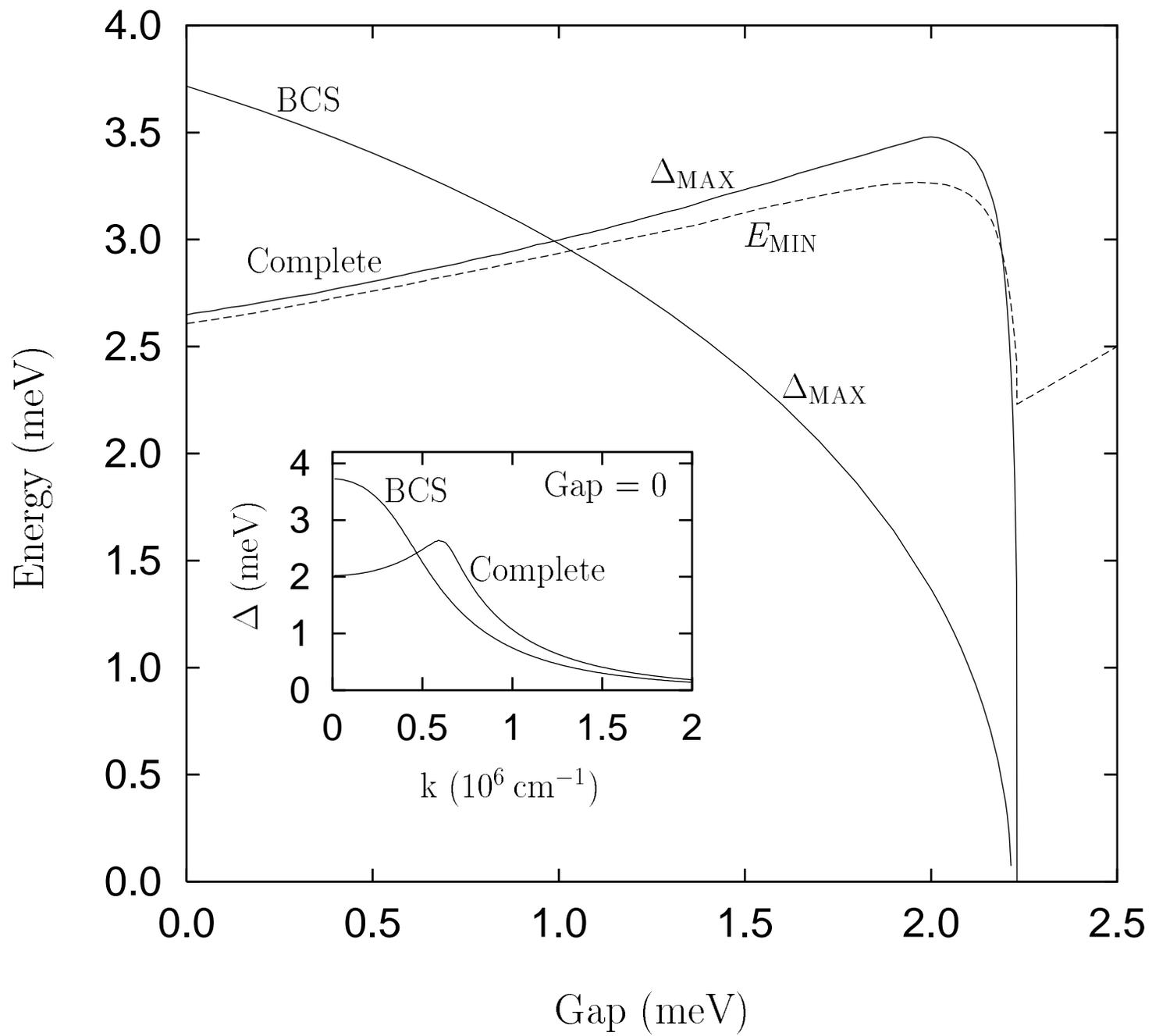
\end{document}